\documentclass[preprint,amsmath,amssymb,prb,floatfix,superscriptaddress]{revtex4}

\usepackage[dvips]{graphicx}
\usepackage{dcolumn}
\usepackage{bm}
\usepackage{mathrsfs}

\begin{document}
\title{Beyond the effective mass approximation: predictive theory of the nonlinear optical response of conduction electrons}
\author{Shukai Yu}
 \affiliation{Department of Physics and Engineering Physics, Tulane University, 6400 Freret St., New Orleans, LA 70118, USA}
\author{Kate H. Heffernan}
 \affiliation{Department of Physics and Engineering Physics, Tulane University, 6400 Freret St., New Orleans, LA 70118, USA}
\author{Diyar Talbayev}
 \email{dtalbayev@gmail.com}
 \affiliation{Department of Physics and Engineering Physics, Tulane University, 6400 Freret St., New Orleans, LA 70118, USA}

\date{\today}

\newcommand{\cm}{\:\mathrm{cm}^{-1}}
\newcommand{\T}{\:\mathrm{T}}
\newcommand{\mc}{\:\mu\mathrm{m}}
\newcommand{\ve}{\varepsilon}
\newcommand{\dg}{^\mathtt{o}}

\begin{abstract}
We present an experimental and computational study of the nonlinear optical response of conduction electrons to intense terahertz (THz) electric field.  Our observations (saturable absorption and an amplitude-dependent group refractive index) can be understood on the qualitative level as the breakdown of the effective mass approximation.  However, a predictive theoretical description of the nonlinearity has been missing.  We propose a model based on the semiclassical electron dynamics, a realistic band structure, and the free electron Drude parameters to accurately calculate the experimental observables in InSb.  Our results open a path to predictive modeling of the conduction-electron optical nonlinearity in semiconductors, metamaterials, as well as high-field effects in THz plasmonics.

\end{abstract}

\maketitle

Studies of the nonlinear optical response of conduction electrons in semiconductors have become possible in the last decade due to the advent of high-field THz sources\cite{hebling:b6,hoffmann:a29,hirori:091106} that opened a new frontier in nonlinear optics.  A rich variety of ultrafast nonlinear phenomena has been reported\cite{danielson:237401,hirori:081305,hebling:035201,hoffmann:151110,hoffmann:231108,kuehn:146602,kuehn:075204,blanchard:107401,junginger:147403,turchinovich:201304,sharma:18016},  many of which result from highly nonequilibrium excited states of electrons induced by the THz field and the breakdown of the effective mass approximation.  While the emerging phenomenology of the nonlinear propagation is understood qualitatively, the quantitative theoretical connection between the observed phenomena and the basic materials' properties has been missing.  The predictive theory of the optical nonlinearity poses an important fundamental question: while the linear propagation is completely described by the Drude dielectric function, what parameters describe the nonlinear propagation?  Which of the Drude parameters can be retained in the description of the nonlinear polarizability, and which ones must be abandoned?  The answer should have broad implications beyond the nonlinear propagation in semiconductors: the nonlinearity is becoming increasingly important in THz metamaterials and plasmonics\cite{liu:345,fan:217404,jeong:171109}, where the propagating fields can be further enhanced via subwavelength confinement.  The explosive growth in both these fields has relied heavily on the predictive modeling of the optical properties, which enables the exploration of the vast materials' parameter space and the tailored design of metamaterials.  Therefore, the ability to model the nonlinear response of conduction electrons would enable the design of nonlinear metamaterials and plasmonic structures, which has remained an uncharted territory.  

In this article, we explore the optical nonlinearity due to conduction electrons in indium antimonide, InSb.  Experimentally, we observe an amplitude-dependent group refractive index (a delay in the arrival time of the THz pulse) and saturable absorption (an increase in transmission at high incident field), which result from the acceleration of conduction electrons to high crystal momenta and energies by the THz electric field.  We propose a model based on the realistic InSb band structure and the semiclassical electron dynamics to account for the measured nonlinearity.  The nonlinear polarizability is computed using the electron density $n$ and scattering rate $\gamma$ determined as Drude parameters from the linear optical properties.  The computational implementation of the model using the finite-difference time-domain (FDTD) method provides a good quantitative agreement with the experiment.  Our main finding is the ability to use the linear Drude parameters, $n$ and $\gamma$, to accurately describe the nonlinear response.  The effective mass approximation is replaced by the realistic band structure.  Thus, we establish the connection between linear and nonlinear THz optical properties and provide a framework for predictive modeling of the nonlinearity in other semiconductors, metamaterials, and plasmonic structures\cite{hebling:035201,hoffmann:151110,blanchard:107401,turchinovich:201304,liu:345,fan:217404,jeong:171109}.

\begin{figure}[ht]
\begin{center}
\includegraphics[width=3in]{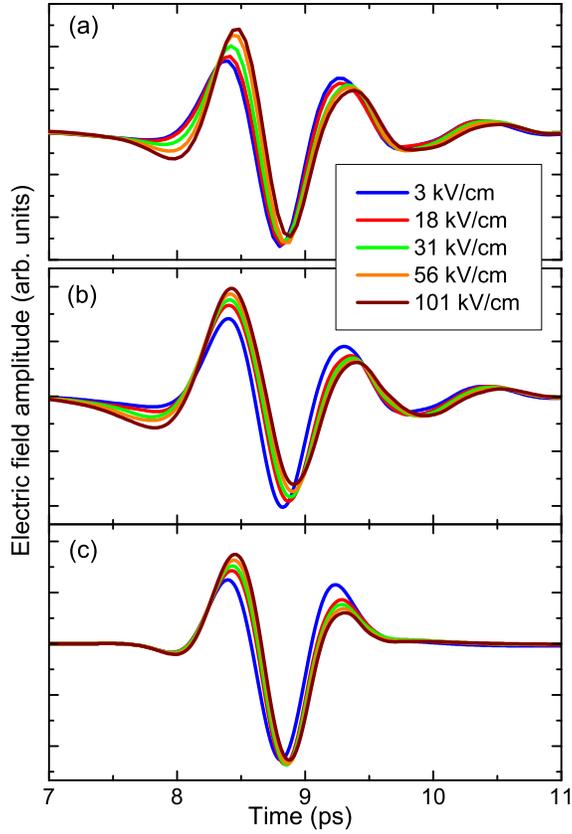}
\caption{\label{fig:1} (a) Experiment: measured evolution of the transmitted THz pulse with increasing peak amplitude of the incident THz field.  The legend box is the same for all three panels and gives the incident peak electric field.  Data in all three panels are normalized to the peak incident field. (b) Simulation: computed evolution of the transmitted realistic source pulse.  (c) Simulation: computed evolution of the transmitted Gaussian source pulse.}
\end{center}
\end{figure}

The experiments were performed using a home built THz spectrometer based on a 1 kHz repetition rate regenerative amplifier\cite{silwal:092116} and THz emission from a LiNbO$_3$ prism with tilted wave front phase matching\cite{hebling:b6,hoffmann:a29,hirori:091106}.  We used electro-optic sampling\cite{planken:313} in ZnTe to estimate the peak THz electric field at the sample to be $\simeq100$ kV/cm.  The sample was mounted on the cold finger of a closed cycle He cryostat and was held at 10 K.  The amplitude of the incident THz field on the sample was controlled by inserting Si attenuators in the THz beam path.  In each measurement, a total of six Si attenuators were used and inserted in the parts of the spectrometer where the THz beam is collimated.  To vary the THz field at the sample position, the attenuators were moved from before to after the sample in the THz beam path.  Each attenuator reduced the peak incident THz field by a factor of 0.56, with an almost flat frequency response.  InSb is a low bandgap semiconductor with a direct bandgap of 0.24 eV at 0 K\cite{littler:986}, a low electron effective mass\cite{goldberg:191} $m^*=0.014m_0$, and a large nonparabolicity of the conduction band\cite{kane:249,cohen:789}.  We used a slightly n-doped (nominally undoped) 0.5 mm thick (100) oriented InSb wafer for these measurements.  The THz electric field was polarized along the $\langle 100 \rangle$ direction in the InSb crystal.

Figure~\ref{fig:1}(a) shows the measured evolution of the transmitted THz pulse with increasing incident peak field.  Two phenomena are apparent in the figure, where the data are normalized to the incident peak amplitude.  First, the THz pulse arrives later at higher incident fields, the amplitude-dependent group delay.  Second, the transmitted peak amplitude increases at higher incident fields, the saturable absorption.  Nonlinear propagation of the THz pulse in InSb was studied in a $z$-scan measurement by Wen et al., who found a drop in transmission at the highest incident field\cite{wen:125203}.  They explain their findings by electron multiplication via impact ionization as the THz field accelerates electrons to energies above the bandgap.  The impact ionization was also observed in subsequent THz pump-probe studies of InSb\cite{hoffmann:161201}.  Direct impact ionization is a very fast process that occurs on $\simeq140$ fs time scale\cite{tanimura:06849,tanimura:045201} with a very low threshold of the incident THz electric field\cite{asmontas:1241} estimated to be $\sim8$ kV/cm.  Why is it not observed in our measurement?  We conjecture that the amount of impact ionization may strongly depend on the initial density of electrons.  Electron multiplication is a cascading process\cite{wen:125203}, as the accelerated electrons create new electrons that are also accelerated and create yet more electrons.  In our sample, the measured electron density is $n=7.3\times10^{13}$ cm$^{-3}$ at 10 K, which is considerably lower than the density cited in the previous studies\cite{wen:125203,hoffmann:161201}.  At sufficiently low initial density, the impact ionization may have no appreciable effect on the propagation of the THz pulse, as evidenced by our experimental data.  

Our data also show little to no evidence of interband electron tunneling due to the intense THz electric field.  Interband electron tunneling was observed under high-field THz excitation in GaAs\cite{kuehn:075204}.  The electron tunneling would lead to the increase in conduction electron density and a drop in high-field THz transmission due to increased THz absorption.  We only observe the transmission increase at the highest THz field, which is completely explained by the nonlinear electron dynamics at fixed electron density, as we will show below.  The negligible electron tunneling rate in our experiments results from a combination of factors.  First, the peak THz field in our measurement is 100 kV/cm, compared to 300 kV/cm reported for the observation of electron tunneling in GaAs\cite{kuehn:075204}, which partially offsets the possible increase in the tunneling probability due to the much lower bandgap in InSb, 0.24 eV, compared to about 1.5 eV in GaAs.  Second, the tunneling rate is proportional to the joint density of states at the edges of the conduction and valence bands, which in turn scales as $(m^*)^{3/2}$.  Due to the much lower electron and hole effective masses in InSb, the joint density of states is more than an order of magnitude lower in InSb than in GaAs.  Finally, the electron tunneling rate due to THz field was reported by Kuehn $et$ $al.$\cite{kuehn:075204} to be proportional to the electron decoherence rate that scales linearly with momentum relaxation rate $\gamma$, electron effective mass, and temperature.  We estimate the electron decoherence rate in our measurement to be two orders of magnitude lower than in GaAs when the electron tunneling was observed\cite{kuehn:075204}.

\begin{figure}[ht]
\begin{center}
\includegraphics[width=3in]{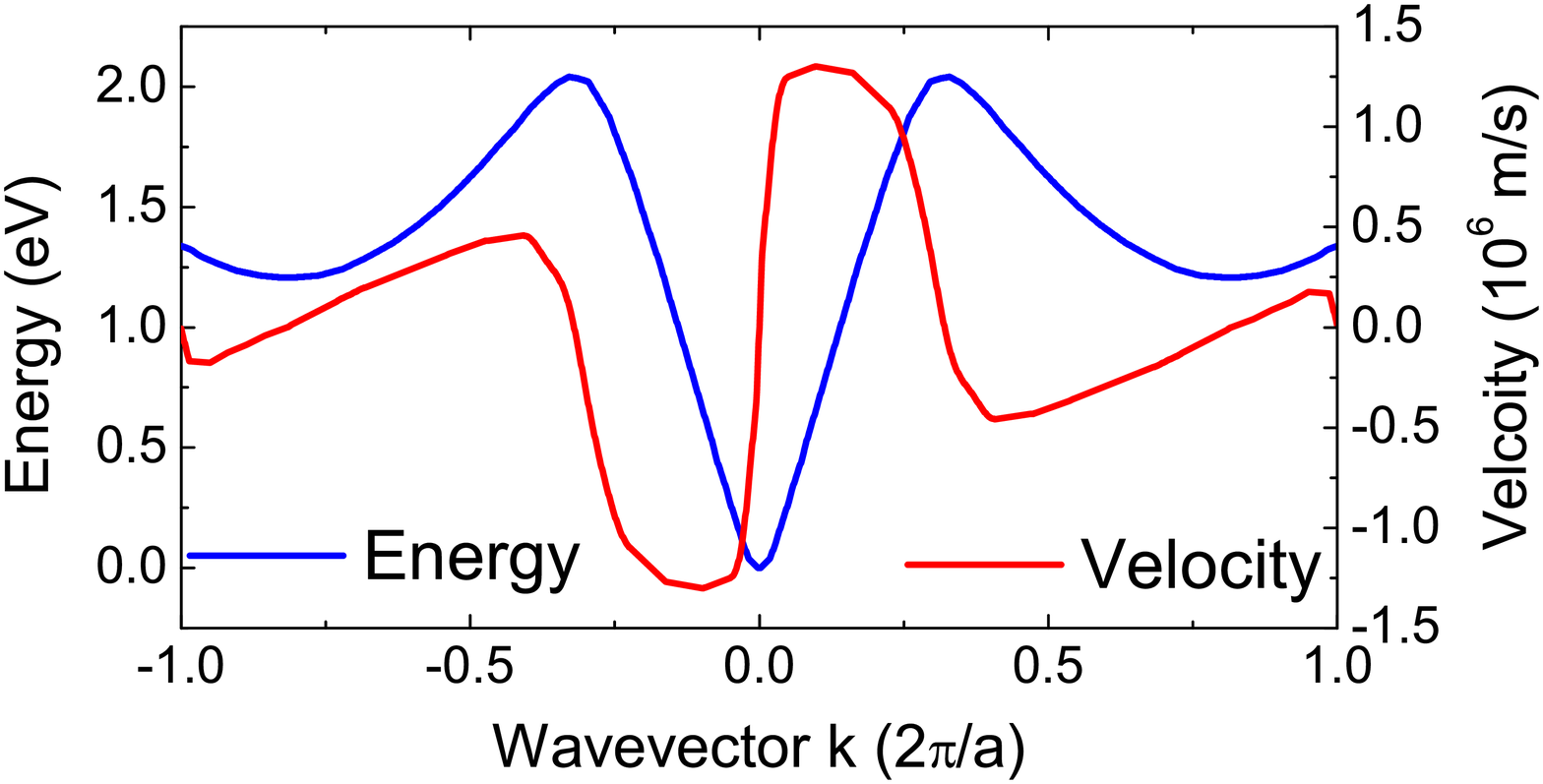}
\caption{\label{fig:bandstr} Conduction band of InSb between $\Gamma$ and X points\cite{kim:035203} and the corresponding electron velocity.  The lattice constant $a=0.64794$ nm\cite{kim:035203}.}
\end{center}
\end{figure}

The observed THz nonlinearity results from the breakdown of the effective mass approximation as the electrons accelerate to energies.  We model the nonlinear propagation computationally using the one-dimensional FDTD method\cite{taflove:fdtd}, which relies on the Yee algorithm\cite{yee:302} to solve for the propagating THz fields $\bm{E}$, $\bm{D}$, and $\bm{H}$ inside InSb.  For the linear propagation, the connection between the fields $\bm{E}$ and $\bm{D}$ is provided by
\begin{equation}
\bm{D}(\omega)=\varepsilon_0\varepsilon(\omega)\bm{E}(\omega)
\label{eq:lin}
\end{equation}
with the Drude dielectric function
$\varepsilon(\omega)=\varepsilon_{\infty} ( 1-\omega_p^2 /(\omega^2+i\omega\gamma))$,
where the $\varepsilon_{\infty}=15.6$ is the background dielectric constant and the plasma frequency $\omega_p^2=ne^2/\epsilon_0\epsilon_{\infty}m^*$ is related to the electron density $n$ and electron effective mass $m^*$.  Equation (\ref{eq:lin}) relies on the effective mass approximation and is no longer useful when the approximation breaks down.  Instead, we compute the electric displacement as $\bm{D}=\varepsilon_0\bm{E}+\bm{P}$, where the polarization $\bm{P}$ consists of linear ($L$) and nonlinear ($NL$) parts: $\bm{P}=\bm{P}^L+\bm{P}^{NL}$.  The linear part is due to the background high-frequency  dielectric susceptibility $\chi_0$: $\bm{P}^L=\varepsilon_0\chi_0\bm{E}$, and the linear part of the displacement becomes $\bm{D}=\varepsilon_0\bm{E}+\varepsilon_0\chi_0\bm{E}=\varepsilon_0\varepsilon_{\infty}\bm{E}$.  We then add the nonlinear part $\bm{P}^{NL}$ to this equation and obtain the relationship 
\begin{equation}
\bm{D}(r)=\varepsilon_0\varepsilon_{\infty}\bm{E}(r)+\bm{P}(r),
\label{eq:dep}
\end{equation}
where we have dropped the $NL$ superscript from $\bm{P}(r)$ that now denotes the nonlinear polarization induced by the electric field $\bm{E}(r)$.  We model the conduction electron response using the semiclassical description\cite{ashcroft:ssp}, in which the velocity of an electron wave packet of wavevector $\bm{k}$ is given by 
\begin{equation}
\bm{v}(\bm{k})=\frac{1}{\hbar}\frac{\partial \mathcal{E}(\bm{k})}{\partial \bm{k}},
\label{eq:bandvel}
\end{equation} 
where $\mathcal{E}(\bm{k})$ is the conduction band energy-momentum dependence.  The electron response to electric field is governed by the equation of motion 
\begin{equation}
\hbar\dot{\bm{k}}+\gamma \hbar \bm{k}=-e\bm{E}(r,t),
\label{eq:eqm}
\end{equation}
where $\gamma$ is the electron scattering rate.  Polarization $\bm{P}$ is computed as $\dot{\bm{P}}=-ne\bm{v}$, where $n$ is the electron density and the velocity $\bm{v}$ must be determined from (\ref{eq:bandvel}) using a realistic band structure of InSb after solving the equation of motion (\ref{eq:eqm}).  Figure~\ref{fig:bandstr} shows the conduction band energy-momentum dependence $\mathcal{E}(\bm{k})$ and the velocity $\bm{v}(\bm{k})$ in InSb used in our computation\cite{kim:035203,malone:105503,mohammad:2935}.

We combine the equations (\ref{eq:dep})-(\ref{eq:eqm}) to obtain a sequence of finite-difference equations that connect the $\bm{E}$ and $\bm{D}$ fields for the Yee algorithm. The THz electric field $\bm{E}$ propagates along the $x$ direction and is polarized along $z$.  The fields $\bm{E}$ and $\bm{D}$ only have $z$ components, as do the polarization $\bm{P}$ and the electron wave vector $\bm{k}$.  The field $\bm{H}$ has only $y$ components.  The FDTD algorithm computes the fields $H$, $D$, and $E$ for the time step $l+1$ by using the field values from the earlier time steps, which are assumed known and stored in computer memory.  We denote the time increment at each step as  $\Delta t$ and write the time point for the $l$-th step as $t^l=l\Delta t$.  We first compute the fields $H$ and $D$ using Maxwell's curl equations and the Yee central differencing in time and space for a nonpermeable medium\cite{taflove:fdtd}.  This step does not yet include the connection to the conduction electron polarizability, which we establish next.  From the equation of motion (\ref{eq:eqm}) we obtain the equation for the wave vector $k$ for the time step $l+1$ at each point on the spatial grid 
\begin{equation}
k^{l+1}=(1-\gamma\Delta t)k^l-(e \Delta t/\hbar)E^l.
\label{eq:fdk}
\end{equation}  
After computing the wave vectors $k^{l+1}$, we determine the velocities $v^{l+1}$ from the realistic band structure (Fig.~\ref{fig:bandstr}) and use them to compute the polarization and the electric field as
\begin{eqnarray}
\label{eq:fdp}
P^{l+1}=P^{l-1}-ne\Delta t (v^{l+1}+v^{l-1}),\\
\label{eq:fde}
E^{l+1}=(D^{l+1}-P^{l+1})/(\varepsilon_0\varepsilon_\infty).
\end{eqnarray}
The Yee algorithm relies on finite-difference equations that are central about the time point $t^l$, and equation (\ref{eq:fdk}) is not.  This is because the electric field value $E^{l+1}$ is not available to compute $k^{l+1}$ at this stage in the algorithm.  To preserve the central difference nature of the algorithm, we compute $k^{l+1}$ again after the field $E^{l+1}$ is known using
\begin{equation}
\label{eq:fdkk}
k^{l+1}=\frac{1-\gamma\Delta t}{1+\gamma\Delta t}k^{l-1}-\frac{e\Delta t}{\hbar(1+\gamma\Delta t)}\left( E^{l+1} + E^{l-1} \right).
\end{equation}
We then repeat the steps (\ref{eq:fdp}) and (\ref{eq:fde}).  This completes the computation of the fields $E$, $D$, and $H$ for the time point $t^{l+1}$.  

We used the above computational model to simulate the nonlinear propagation of the THz pulse in InSb.  We determined the electron density $n$ and the scattering rate $\gamma$ from the linear THz spectroscopic measurement with very low incident electric field: $n=7.3\times10^{13}$ cm$^{-3}$ and $\gamma=0.5$ THz.  The time increment was set to $\Delta t=2.08$ fs and the space increment $\Delta x$ was set to 5 $\mu$m in vacuum and 1.25 $\mu$m inside InSb.  We used two different incident source pulses: one was the realistic THz pulse recorded as the free-space reference in our measurement; the other was a Gaussian pulse given by $E(t)=E_0\exp\left[ -t^2/\tau_d^2\right] \sin\left[ 2\pi f_0t \right] $ with $\tau_d=0.5$ ps and $f_0$=1 THz.  The results of the computed nonlinear propagation of these THz pulses through a 0.5 mm InSb layer are shown in Figs.~\ref{fig:1}(b,c).  Comparison with the experimental data (Fig~\ref{fig:1}(a)) shows that the computation reproduces very well the major features of the nonlinear THz propagation - the group delay, the increase in the peak transmitted THz field, and the overall evolution of the pulse shape as the incident field gets stronger.

\begin{figure}[ht]
\begin{center}
\includegraphics[width=3in]{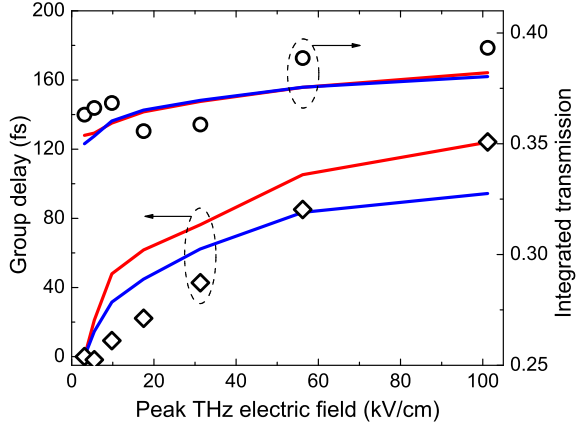}
\caption{\label{fig:gdtm} Measured frequency-integrated transmission (circles) and group delay (diamonds) compared to the results of the simulated nonlinear THz propagation of the realistic (red line) and Gaussian (blue line) source pulses.}
\end{center}
\end{figure}

For a more quantitative comparison between the experiment and computation, we measured the change in the arrival time of the THz pulse (the group delay) with increasing peak THz field.  We define the group delay as $\Delta t=(c\Delta\phi)/(2\pi d)$, where where $d$ is the sample thickness, $c$ is the speed of light, and $\Delta\phi$ is the average phase difference between high- and low-amplitude pulses transmitted by the sample in the frequency domain:
\begin{equation}
\Delta\phi=\left\langle \arg\left[ \frac{E_{high}(\omega)}{E_{low}(\omega)}\right] \right\rangle.
\end{equation}
Here, we average over the full frequency content of the THz pulse.  We also quantified the frequency-integrated transmission as\cite{sharma:18016}
\begin{equation}
T=\frac{\int E^2_{sam}(t) dt}{\int E^2_{ref}(t) dt},
\label{eq:intt}
\end{equation}
where $E_{sam}$ and $E_{ref}$ are the time-domain electric fields transmitted by the InSb sample and the free space reference.  Figure~\ref{fig:gdtm} shows the experimental and computational group delay and integrated transmission.  Our computational model describes very well the overall behavior of both parameters and provides a good quantitative agreement at the highest experimentally available peak THz fields.  We emphasize that there are no free (fitting) parameters in the computational results of Fig.~\ref{fig:gdtm}, as the electron density $n$ and the scattering rate $\gamma$ were fixed to the values obtained from the linear spectroscopic measurement.  The only variable is the strength of the incident THz electric field.

\begin{figure}[ht]
\begin{center}
\includegraphics[width=3in]{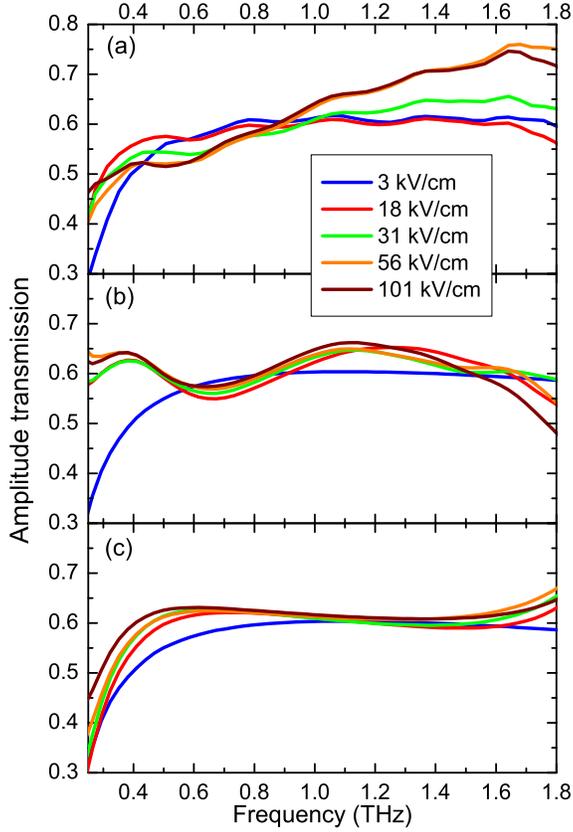}
\caption{\label{fig:freqdomtrans} (a) Experiment: measured amplitude transmission as a function of the incident peak THz amplitude.  The legend box is the same for all three panels and gives the incident peak electric field.  (b) Simulation: computed amplitude transmission with the realistic source pulse.  (c) Simulation: computed amplitude transmission with the Gaussian source pulse.}
\end{center}
\end{figure}

We now look at the nonlinear amplitude transmission in the frequency domain (Fig.~\ref{fig:freqdomtrans}).  The transmission measured in the linear regime at the lowest incident field (the blue line in all three panels in Fig.~\ref{fig:freqdomtrans}) is reproduced very well by the model.  The linear transmission displays the characteristic dip below $\sim 0.5$ THz that corresponds to the Drude response of free electrons.  At high incident field in the nonlinear regime, we find differences between the experiment and the computational results.  We find notable differences even between the realistic source and the Gaussian source computational results (Figs.~\ref{fig:freqdomtrans}(b,c)).  Despite the differences, the model captures well the main features of the nonlinear response, such as the increased transmission below 0.5 THz due to the saturation of the free electron absorption.  The model also reproduces the nonmonotonic frequency dependence of the nonlinear transmission, which is apparent in Figs.~\ref{fig:freqdomtrans}(a,b): the transmission increases below 0.5 THz, decreases around 0.7 THz, and increases again above 1 THz.  Another detail found in both the experiment and model is the nonmonotonic dependence of the nonlinear transmission on the incident THz amplitude at a specific frequency: for example, at 1.6 THz, the transmission first goes up and then goes down as the incident amplitude increases.  

The difference between the nonlinear transmission computed using the realistic and Gaussian source pulses (Figs.~\ref{fig:freqdomtrans}(b,c)) does not come as a surprise.  The realistic and Gaussian source pulses differ in their time domain shape and frequency content.  The result of their nonlinear propagation cannot be obtained by applying the same transfer function to both pulses, as each Fourier frequency component will be enhanced or suppressed differently, depending on its amplitude and on the amplitudes of other Fourier components.  Therefore, the different transmission for different time-domain inputs is a hallmark of nonlinear propagation.  The same reasoning explains the difference in the measured and computed nonlinear transmission in Figs.~\ref{fig:freqdomtrans}(a,b).  Even though the realistic source pulse in the model is the measured reference THz pulse, its measurement includes the response function of the THz receiver.  The THz pulse that interacts nonlinearly with the InSb sample does not include the receiver response function and could have a different time domain shape from the measured reference pulse.  This difference causes no adverse effects in a linear sample-reference spectroscopic measurement, as the receiver response function cancels out when the linear transmission is computed.  In our nonlinear case, this difference explains the discrepancy in transmission between the measurement and the realistic source model (Figs.~\ref{fig:freqdomtrans}(a,b)). 

The significance of our results is that a simple semiclassical model of electron dynamics, with the realistic band structure and the linear Drude parameters $n$ and $\gamma$, provides a good quantitative description of nonlinear THz propagation.  This is not an \textit{a priori} expected conclusion.  When electrons are accelerated to high energies by the THz field, strong intervalley scattering is expected\cite{hoffmann:151110,hebling:035201}, potentially resulting in increased scattering rates and nontrivial electron distributions in the Brillouin zone.  Specifically in InSb\cite{tanimura:045201}, the $\Gamma-$L valley scattering takes place in about 40 fs.  Nonetheless, we use the fixed scattering rate $\gamma$ and the average electron wavevector $\bm{k}$ to successfully describe the nonlinear THz optical properties.  The saturable absorption and the increased group refractive index have also been reported in other semiconductors\cite{hoffmann:151110,turchinovich:201304}, where they exhibit very similar dependence on the strength of the incident THz field.  This suggests that the validity of the proposed model should extend beyond the presented phenomenology of InSb.  It is easily applied to other semiconductors by using the appropriate band structure.

The importance of our model extends beyond the presented one-dimensional propagation in a uniform medium.  The computational model is easily incorporated in the FDTD descriptions of THz nonlinearity in two- and three-dimensional situations, e.g., when the semiconductor forms a part of a metamaterial\cite{fan:217404,jeong:171109}.  While we use here the average electron $\bm{k}$ vector for simplicity, the method allows straightforward extensions to Monte Carlo-type descriptions of electron dynamics that make use of a distribution of electronic states and/or more sophisticated scattering models\cite{asmontas:1241}.  However, the implementations of more complex models of electron dynamics place much higher demands on computational resources (computer memory and processor time), especially in two- and three-dimensional geometries.  This emphasizes the value of the presented much simpler model that quantitatively describes the THz optical nonlinearity.  

To conclude, we have studied the nonlinear optical properties of conduction electrons in InSb.  Despite the potential complexity of electron dynamics and scattering at high energies, the nonlinear polarization model based on the realistic band structure, semiclassical dynamics, and the measured Drude parameters is sufficient to compute the observed nonlinear properties.  Our results open a path to the unified and predictive description of the THz optical nonlinearity across many semiconductors, metamaterials, and plasmonic structures.

This work was supported by the Louisiana Board of Regents contracts LEQSF(2012-15)-RD-A-23 and LEQSF-EPS(2014)-PFUND-378, and by the NSF award number DMR-1554866.  


\end{document}